\documentclass[twocolumn]{emulateapj}
\usepackage{apjfonts}
\usepackage{natbib}
\def\la{\mathrel{\hbox{\rlap{\hbox{\lower4pt\hbox{$\sim$}}}\hbox{$<$}}}}
\def\ga{\mathrel{\hbox{\rlap{\hbox{\lower4pt\hbox{$\sim$}}}\hbox{$>$}}}}
\def\arcdeg{\hbox{$^\circ$}}
\def\arcmin{\hbox{$^\prime$}}

\shorttitle{Non-Thermal Emission in the Coma Cluster}
\shortauthors{Wik et al.}

\begin{document}

\title{A {\it Suzaku} Search for Non-thermal Emission at Hard X-ray Energies 
in the Coma Cluster}

\author{Daniel R. Wik\altaffilmark{1}, Craig L. Sarazin\altaffilmark{1},
Alexis Finoguenov\altaffilmark{2,3}, Kyoko Matsushita\altaffilmark{4},
Kazuhiro Nakazawa\altaffilmark{5}, Tracy E. Clarke\altaffilmark{6,7}}

\altaffiltext{1}{Department of Astronomy, University of Virginia, 
P. O. Box 400325
Charlottesville, VA 22904-4325; 
drw2x@virginia.edu}
\altaffiltext{2}{Max Planck Institute for Extraterrestrial Physics}
\altaffiltext{3}{Center for Space Science Technology, University of 
Maryland Baltimore County}
\altaffiltext{4}{Department of Physics, Tokyo University of Science}
\altaffiltext{5}{Physics Department, University of Tokyo}
\altaffiltext{6}{Naval Research Laboratory}
\altaffiltext{7}{Interferometrics Inc.}

\begin{abstract}
The brightest cluster radio halo known resides in the Coma cluster of galaxies.
The relativistic electrons producing this diffuse synchrotron emission
should also produce inverse Compton emission that becomes competitive
with thermal emission from the ICM at hard X-ray energies.
Thus far, claimed detections of this emission in Coma
are controversial
\citep[e.g.,][]{FOB+04, RM04}.
We present a {\it Suzaku} HXD-PIN observation of the Coma cluster
in order to nail down its non-thermal hard X-ray content.
The contribution of thermal emission to the HXD-PIN spectrum is constrained
by simultaneously fitting thermal and non-thermal models to it and 
a spatially equivalent spectrum derived from an {\it XMM-Newton} mosaic 
of the Coma field \citep{SFM+04}.
We fail to find statistically significant evidence for non-thermal 
emission in the spectra,
which are better described by only a single or multi-temperature model
for the ICM.
Including systematic uncertainties, we derive a 90\% upper limit on the
flux of non-thermal emission of $6.0\times10^{-12}$ erg s$^{-1}$ cm$^{-2}$
(20-80 keV, for $\Gamma=2.0$),
which implies a lower limit on the cluster-averaged magnetic field of
$B>0.15 \, \mu$G.
Our flux upper limit is $2.5\times$ lower than the detected non-thermal
flux from {\it RXTE} \citep{RG02} and {\it BeppoSAX} \citep{FOB+04}.
However, if the non-thermal hard X-ray emission in Coma is more spatially
extended then the observed radio halo, the {\it Suzaku} HXD-PIN may 
miss some fraction of the emission.
A detailed investigation indicates that $\sim$50--67\% of the emission
might go undetected, which could make our limit
consistent with \citet{RG02} and \citet{FOB+04}.
The thermal interpretation of the hard Coma spectrum is consistent with
recent analyses of {\it INTEGRAL} \citep{ENC+07} and {\it Swift}
\citep{Oka+08, Aje+09} data.
\end{abstract}


\section{Introduction} \label{sec:intro}

In the hierarchical scenario of cosmic structure formation, clusters
of galaxies form at late times through mergers between subclusters 
and through the accretion of galaxies and galaxy groups.
The distribution of their massive halos in space and time depend
sensitively on the underlying cosmology, and much effort has been
made to connect observable properties of the gas to the total cluster
mass in order to constrain cosmological parameters 
\citep[e.g.,][and references therein]{MAE+08}.
However, merger processes are known to significantly disrupt the thermal
gas \citep[e.g.][]{RS01, RT02},
typically biasing inferred masses and the 
resultant cosmological parameter estimates \citep{RSR02, WSR+08}.
Merger-induced shocks and turbulence, besides heating the gas, are thought
to also re-accelerate relativistic particles present in the intracluster
medium (ICM) \citep{Sar99, BB05}.
Non-thermal electrons, observed via diffuse, radio synchrotron emission,
have been detected in over 50 clusters, 
all of them undergoing mergers \citep{Buo01, SBR+01}.
If the energy in a relativistic phase of the ICM is large enough to add
pressure support to the thermal gas, even transiently, the ability
to derive masses and therefore use clusters as cosmological probes may
be compromised \citep{SOH+08}.
An assessment of the relativistic contribution to the energy budget
of clusters is necessary to fully characterize the state of the ICM.

Diffuse, cluster-wide synchrotron radio emission, called radio halos or
relics depending on their morphology, imply that both magnetic fields and
relativistic electron populations are present on large scales.
The total luminosity of a synchrotron-emitting electron is given by
\begin{equation} \label{eq:lsync}
L_R = \frac{4}{3}\sigma_Tc\gamma^2\epsilon_{B}
\, ,
\end{equation}
where $\sigma_T$ is the Thomson cross-section, $c$ is the speed of light,
$\gamma$ is the Lorentz factor of the electron, and $\epsilon_B=B^2/8\pi$ is the
energy density of the magnetic field.
For many such electrons, the value of $L_R$ depends both on the number of
electrons and on $B$ and cannot independently determine either.
However, these same electrons will up-scatter cosmic microwave
background (CMB) photons through inverse Compton (IC) interactions, which have
a luminosity $L_X$ equivalent in form to equation~(\ref{eq:lsync}) but with
$\epsilon_B$ replaced by the energy density of the CMB.
Since both luminosities are proportional to the number of electrons,
their ratio gives the volume-averaged magnetic field,
\begin{equation} \label{eq:synicratio}
\frac{L_R}{L_X} = \frac{B^2/8\pi}{aT_{CMB}^4}
\, ,
\end{equation}
where $a$ is the radiation constant and $T_{CMB}$ is the temperature of the
CMB.
The IC radiation should be observable at hard X-ray energies \citep{Rep77}.
Thus far, IC emission has only been detected at low significance
\citep{NOB+04} or, in one case, in a cluster with uncertain radio emission
(\citealt{EPP+08}; but see also \citealt{Aje+09} and \citealt{Fuj+08}).
The measurement of an IC flux from a synchrotron source directly leads
to a simultaneous determination of the average value of $B$ and
the relativistic electron density \citep{HR74, Sar88}.
Therefore searches for IC emission coincident with a radio
halo or relic are an excellent way to constrain the
contribution of relativistic materials in clusters.

The first, and brightest, radio halo was discovered by \citet{Wil70} in
the Coma cluster, and its radio properties have perhaps been the best
studied \citep[e.g.][]{GFV+93, DRL+97, TKW03}.
Coma has been observed by all the major observatories with hard X-ray
capabilities \citep{RUG94, HBS+93, Baz+90, HM86},
and more recently non-thermal detections have been
claimed by \citet{RG02} with {\it RXTE} and by
\citet{FDF+99, FOB+04} with {\it BeppoSAX}, though
the latter detection is controversial \citep{RM04, FLO07}.
Due to the large field of view (FOV) of these non-imaging instruments
and the simple characterization of the thermal gas, the source
of this emission remains uncertain.
Even more recently, long ($\sim 1$ Msec) observations with {\it INTEGRAL}
have imaged extended diffuse hard X-ray emission from Coma, though it
was found to be completely consistent with thermal emission
\citep{RBP+06, ENC+07, LVC+08}.

In this study, we present a {\it Suzaku} HXD-PIN observation of the Coma
cluster in an effort to detect non-thermal emission associated with the 
radio halo and potentially confirm the {\it RXTE} and {\it BeppoSAX} detections.
The HXD-PIN instrument has a non-imaging collimator like those on-board
{\it RXTE} and {\it BeppoSAX}, but with a FOV about a quarter as large, which
reduces the possible contamination from hard point sources \citep{Tak+07}.
Also, the {\it Suzaku} particle background is $\sim5\times$ lower
than the backgrounds of either {\it RXTE} or {\it BeppoSAX} \citep{Mit+07}.
In order to clearly distinguish the thermal and non-thermal emission 
visible within the PIN, the hard {\it Suzaku} PIN spectrum is jointly
fit with a spatially equivalent {\it XMM-Newton} EPIC-pn spectrum.
The {\it XMM} spectrum, at lower energies and completely dominated by thermal
emission, allows Coma's thermal and non-thermal properties to be
simultaneously determined.
The {\it XMM} and HXD-PIN observations are reported in \S~\ref{sec:obs}
and the extraction of the resulting spectra is discussed in
\S~\ref{sec:specextract}.
Fits to the joint spectra are described in \S~\ref{sec:fits}.
In \S~\ref{sec:disc}, we discuss the implications of our results for
the nature of the hard X-ray emission from the Coma cluster.
We assume a flat cosmology with $\Omega_M = 0.23$ and $H_0 = 72$ km/s/Mpc
and a luminosity distance to Coma of 98.4 Mpc.
Unless otherwise stated, all uncertainties are given at the 90\% 
confidence level.

\section{Observations} \label{sec:obs}

The {\it Suzaku} observation was undertaken as part of AO-1 from
2006 May 31 through June 4,
soon after 16 of the 64 PIN diode bias voltages were lowered
from 500V to 400V, but before an additional 16 diodes were similarly lowered.
We analyze Version 2 of the pre-processed data (PROCVER 2.0.6.13), which
allows for the diode bias drop, with HEAsoft 6.4.0 and XSPEC 12.4.0w.
For the HXD-PIN instrument, the standard data selection criteria are applied 
to extract the source spectrum, and the same criteria are used to select
times for the modeled non-X-ray background (NXB) spectrum.
Specifically, we select observing times when the geomagnetic cut-off 
rigidity is above a critical value (COR $> 6$ GV), when the satellite 
is not within the South
Atlantic Anomaly (SAA\_HXD $= 0$) or has just left it
(T\_SAA\_HXD $> 500$ s),
and when {\it Suzaku} is pointed above and at least 5 degrees away from the 
Earth's horizon (ELV $> 5\arcdeg$).
The strength of the NXB is known to be roughly inversely proportional to
the value of the COR and to be elevated inside the SAA, gradually decaying
to typical levels after SAA passage.
These criteria ensure that the low NXB of the HXD is minimized and can be well
characterized, which is necessary if it is to be accurately modeled.
After event selection, the PIN exposure time is reduced from 166.2 ks to
156.1 ks after dead-time correction.
The HXD-GSO spectrum is found to be consistent with the background,
so we do not consider it further here.
We use the {\it Suzaku} CCD data from the XIS0 chip to check the cross-calibration of
{\it Suzaku} and {\it XMM-Newton}.
Standard event selection was applied to the XIS0 data, leading to an exposure time
of 178.7 ks.

The mosaic {\it XMM-Newton} observations of the Coma cluster, including
14 separate pointings, were done as a part of an
instrument performance verification program, a complete log of which is
presented in \citet{SFM+04}.
The initial observations were undertaken by and first reported in 
\citet{Bri+01}.
Seven new observations, aimed at resolving the temperature structure of the 
Coma center, have also been performed (PI P. Schuecker).
However, high solar activity during the exposures resulted in a
high detector background above 2 keV, making these observations less suitable
for our purposes, and therefore we use only the observations reported
in Section~4 and Table~2 of \citet{SFM+04}.
We choose only to include the EPIC-pn data from {\it XMM} in our analysis.
Because these observations were made early in the mission, they cannot be 
processed with the standard software, though the EPIC-pn data have 
undergone in-house processing.
Also, its effective area at high energies is higher than for the EPIC-MOS
detector, making it the more suitable instrument.
The benefit of including the EPIC-MOS data is unclear, 
due to the addition of cross-calibration errors and given the already 
high signal-to-noise of the EPIC-pn data.

\section{Extraction and Construction of Spectra} \label{sec:specextract}

To produce complementary spectra from the 
{\it XMM} EPIC-pn and {\it Suzaku} HXD-PIN 
data that can be simultaneously fit, the background and responses
of both instruments must be carefully considered to minimize systematic
uncertainties.
The expected non-thermal signal is near the limit of the PIN sensitivity,
and a robust characterization of this emission particularly depends on the 
accuracy of the PIN background and {\it XMM}-{\it Suzaku} cross-normalization.

\subsection{HXD-PIN Spectrum and Non-X-ray Background} \label{sec:specpin}

As the HXD is a non-imaging instrument, we simply extract the PIN 
spectrum from the selected events and group the spectral bins so that
each bin contains at least 30 counts to ensure
that Gaussian statistics and $\chi^2$ fitting are valid.
The response matrix is provided in the
{\it Suzaku} CALDB\footnote{http://suzaku.gsfc.nasa.gov/docs/heasarc/caldb/suzaku/}
for Version 2 data products, and we use ae\_hxd\_pinhxnome2\_20080129.rsp
for all source components other than the cosmic X-ray background (CXB), 
for which 
\linebreak ae\_hxd\_pinflate2\_20080129.rsp is used.

\begin{figure}
\plotone{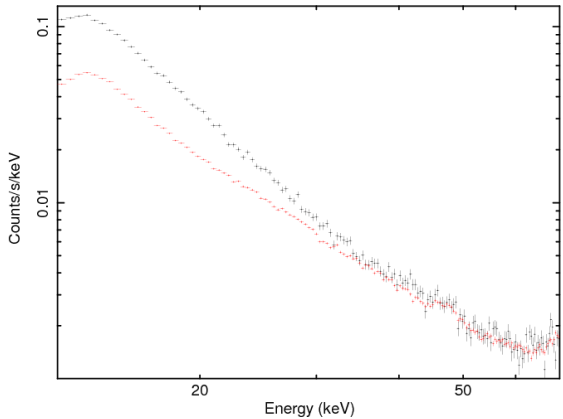}
\caption{{\it Suzaku} HXD-PIN NXB model spectrum (red data points) compared to the
Coma cluster data (uncorrected for background).
Note that at energies above 45 keV, the NXB dominates the data and that
deviations of the data above the NXB are confined to individual channels
that are simply statistical fluctuations or are otherwise imperfectly 
characterized by the NXB model.
\label{fig:nxb_data}}
\end{figure}

\begin{figure}
\plotone{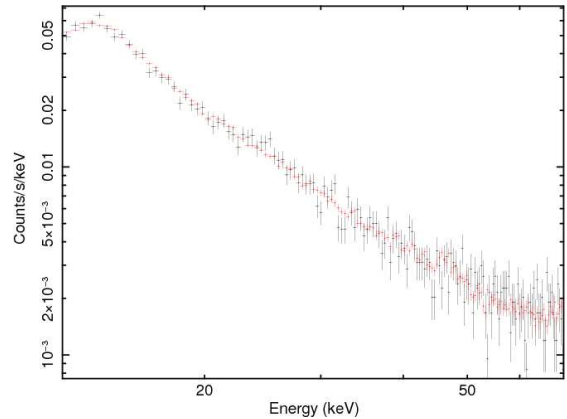}
\caption{{\it Suzaku} HXD-PIN Earth-occulted data (black data points) compared to the
NXB model spectrum for the same time periods of Earth-occultation
(red data points).
During Earth-occultation, the only events should be due to the NXB.
Note that the NXB agrees well with the normalization and shape of the
Earth-occulted data at all energies $\ga$12 keV.
(Only this range is used in the spectral fits for Coma.)
\label{fig:nxb_earth}}
\end{figure}

The non-X-ray background for a PIN observation is most accurately 
obtained from a model, as opposed to a comparable blank field 
observation.
This method is motivated by the strong dependence of the background
count rate and spectral shape on the value of the geomagnetic cut-off 
rigidity (COR) and 
the time since the passage of the satellite through the South Atlantic 
Anomaly (SAA), quantities which vary and have a unique distribution
for every observation.
The model matches the distribution of the COR and other parameters of the
observed data.
A model is also required because there is no concurrent measurement of the
NXB, such as by nodding between Coma and a blank field.
We use the so-called ``bgd\_d" model for Version 2 processed data, which
makes use of HXD-GSO information as well as the COR and SAA values.
This NXB model is shown with the PIN data spectrum, uncorrected for the
background, in Figure~\ref{fig:nxb_data}.
While the shape of the NXB is generally well reproduced, the success 
of the model in determining its overall normalization is $\pm 2.3\%$ from 
15-40 keV and $\pm 4\%$ from 40-70 keV (Mizuno et al., {\it Suzaku} Memo
2008-03\footnote{http://www.astro.isas.ac.jp/suzaku/doc/suzakumemo/suzakumemo-2008-03.pdf}).
These estimates of the systematic error in the NXB are extrapolated to the
90\% confidence interval from the $1\sigma$ values derived using 
Earth-occulted data in {\it Suzaku} Memo 2008-03.
We adopt these values (2.3\% from 12-40 keV and 4\% from 40-70 keV) as 
our estimate of the 90\% systematic error in the PIN NXB.
To confirm the accuracy of the model background, we extracted
events for both the data and model for times when the Earth occults
the PIN FOV (ELV $< -5\arcdeg$, all other selection criteria unchanged);
the resulting spectra are shown in Figure~\ref{fig:nxb_earth}.
The fractional difference between the model NXB count rate and the 
Earth-occulted data is ($0.2 \pm 1.0$)\% over the energies 12-40 keV and
($-2.0 \pm 2.6$)\% from 40-70 keV (1$\sigma$ errors).
Over the whole range considered, the fractional difference in count rates
is extremely small: ($0.005 \pm 0.9$)\%.
Because of the excellent agreement, we do not adjust the level of the
background as proposed in Ishida et al., {\it Suzaku} Memo 2007-10\footnote{http://www.astro.isas.ac.jp/suzaku/doc/suzakumemo/suzakumemo-2007-10.pdf}.

The estimate of the systematic error adopted here comes from an analysis
of Earth-occulted data, which is the same data used to generate
models of the NXB.
However, it is possible that a systematic effect could be present during
observations of the sky that would not exist during Earth-occulted observations,
and so it would not be included in the NXB model or the estimate of the
systematic uncertainty.
Mizuno et al. ({\it Suzaku} Memo 2008-03) attempt to test for this possibility with 
``blank sky" observations and find a larger effective systematic uncertainty.
It is clear that some part of this uncertainty is due to the fact that the
fields aren't entirely ``blank" and that the source flux
will vary field-to-field.
Here we refer to relatively bright sources not considered to be part of the
more uniform CXB, which has a variance based on the XIS 
sensitivity to point sources that can be taken into account.
When considering many observations of one region on the sky, so that the 
contamination from sources will vary less, the standard deviation
drops from 5.8\% derived from many fields to 5.0\%, both of which includes
a statistical error of about 3.3\%.
While an additional systematic uncertainty, only in effect when observing the
sky, cannot be ruled out, this drop suggests that systematic error 
estimates derived from sky observations are somewhat conservative.
Because the contribution of contaminating sources to systematic error
estimates is thus far not well-characterized, we use the value derived
from Earth-occulted observations throughout to avoid overestimating this error.
We rely on the assumption that a full accounting of contaminating sources
would lead to a systematic error estimate similar to our adopted value.
However, using the sample of 10 ks exposures of all blank sky observations
leads to an estimate of the NXB systematic error of 4\%, after subtracting the
statistical error and the expected CXB fluctuation \citep[see e.g.][]{Nak+08}.
We consider the effect of raising the systematic error to this higher value
(for $E<40$ keV) in \S~\ref{sec:syst}; our results and conclusions remain
qualitatively unchanged.

\subsection{{\it XMM} EPIC-pn Spectrum} \label{sec:specxmm}

To constrain the thermal contribution to the PIN hard X-ray spectrum,
it is very helpful to have a spectrum for the same
region covered by the HXD-PIN FOV but extending to lower energies where the
thermal emission is completely dominant.
This low energy spectrum acts as a lever arm on the thermal continuum
so that the properties of the thermal gas can be extracted simultaneously
with a potential non-thermal component.
Because the ICM of Coma is not isothermal and its projected temperature varies
across the cluster, a complimentary spectrum at softer energies must
follow the spatial sensitivity of {\it Suzaku}'s HXD.
Otherwise, localized regions of even slightly hotter gas could mimic
the emission of a non-thermal source at hard energies.
Since the HXD is made up of 64 individual collimators with optical axes 
generally aligned to within $4\arcmin$ of each other, we approximate the
PIN spatial response as a single perfect collimator with a total square
FOV of $D=65\farcm5$ on a side,
\begin{equation} \label{eq:response}
R_{\rm coll} = (D/2-\theta_x)(D/2-\theta_y)/(D/2)^2
\, ,
\end{equation}
where $R_{\rm coll}$ is the fraction of the flux detected at angles of
($\theta_x$, $\theta_y$) from the optical axis along the PIN detector
axes,
relative to a point source located at the center of the HXD FOV
($\theta_x=\theta_y=0$).
We have verified that this model fits the spatial response of the PIN very well.
The complimentary {\it XMM} spectrum is constructed based on this
spatial vignetting of the PIN, which is reasonable for our energy
range of interest ($< 70$ keV).

\begin{figure}
\plotone{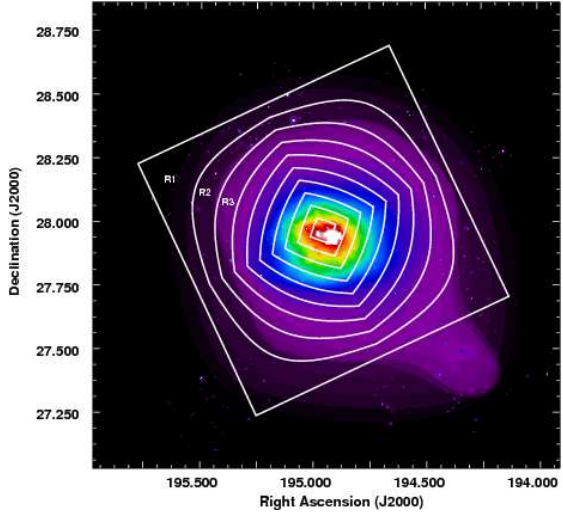}
\caption{{\it XMM} EPIC-pn 2--7.5 keV wavelet smoothed X-ray surface brightness
image \citep{SFM+04} with contours of
constant {\it Suzaku} HXD-PIN effective area overlaid.
The contours are spaced at 10\% intervals running from 0\% to 90\% of the 
effective area for a point source located at the instrument center.
In between the contours, the regions labeled R1, R2, etc., correspond
to those in Table~\ref{tab:reg} and in the text.
\label{fig:xmmim}}
\end{figure}

In order to build a spectrum that reflects the PIN vignetting with good 
statistics, we extract spectra from 10 regions
of roughly equal effective area, as shown in Figure~\ref{fig:xmmim}.
The boundaries of the regions are spaced at intervals of 10\% of the
PIN sensitivity to a central point source.
Because the solid angle subtended by a region increases with its
distance from the cluster center, it turns out that the count rates of each of
these {\it XMM} spectra are comparable.
The same response matrix is used for all spectra, epn\_ef20\_sdY7\_medium.rsp,
and the auxiliary response files (arfs) for each region are generated in the 
standard way \citep{LFS+03}.
The background spectrum is derived from the datasets compiled by
\citet{RP03}, to which we also apply consistent flare cleaning criteria.
Before summing these spectra, weighted by the average PIN sensitivity 
within each region, we scale the arfs so they all agree with the central region
(R10) arf at 5 keV, while also scaling the exposure times so the flux remains
unchanged.
Similarly normalized arfs are required to ensure that the weighted and summed arf
will properly represent the response of the final summed spectrum.

We now describe the procedure for creating the summed observed rate
spectrum, summed background spectrum, and corresponding response.
Let $O^i_j$ be the observed rate spectrum in spectral channel $j$ for
region $i$, and let $B^i_j$ be the corresponding
background spectrum.
We constructed the weighted sums
\begin{equation} \label{eq:xmmsum}
\overline O_j = \sum_{i=1}^{10} w^i O^i_j
\, ,
\end{equation}
\begin{equation} \label{eq:xmmbgd}
\overline B_j = \sum_{i=1}^{10} w^i B^i_j
\, ,
\end{equation}
where $w^i$ is the weight of region $i$ in the PIN spectrum, based on
the average value of 
equation~(\ref{eq:response}) inside the region (Table~\ref{tab:reg}),
normalized by the PIN nominal central point source sensitivity.
Let $R^i_{jk}$ be the response matrix for region $i$, defined such that
given a number flux $M^i_k$ of photons per unit area and time at
Earth in spectral channel $k$,
$R^i_{jk} M^i_k$ is the number of events per second observed in spectral
channel $j$.
In the nomenclature of X-ray spectral analysis, $R^i_{jk}$ is the ``rsp'' file
for region $i$.
The corresponding average response matrix, $\overline R_{jk}$, must be weighted both
by the PIN effective area for the regions (the weights $w_i$) and by
the number of {\it XMM} photons incident on each area.
To include the weighting by the incident flux on each area, 
we formally assume that to first order the spectra in all 10 regions are
described by models $M^i_k$ that have identical spectral shapes given by
$m_k$, but differing normalizations $N^i$.
That is, $M^i_k = N^i m_k$.
In our case, we take $m_k$ as an APEC model with $T=8.2$ keV and abundance
relative to solar of 0.24,
which is the best fit model to the {\it XMM-Newton} EPIC-pn
spectrum for the entire {\it Suzaku} PIN FOV.
While the temperatures from the outer 5 regions (R5--R1, Table~\ref{tab:reg})
are somewhat inconsistent with our fiducial $T$, the central regions
are weighted more strongly and so it is more important to accurately
match their spectral shape than that of the spectra from R5--R1.
The average of the temperatures from R10--R1 is in good
agreement with previous measurements of Coma's global temperature
\citep{Arn+01,WYF+99}.
Also, temperatures in R10, R9, and R8 are just consistent at the 90\% level
with continuum and Fe line ratio fits to the XIS data extracted from those
regions, using the method described in \citet{Sat+08}.
The models $M^i_k$ now differ only in overall flux, given by the APEC
normalization $N^i$, so each observed spectrum can be described as
\begin{equation} \label{eq:regfit}
O^i_j - B^i_j = \sum_k R^i_{jk} M^i_k = N^i \sum_k R^i_{jk} m_k
\, .
\end{equation}
Similarly, we define $\overline R_{jk}$ as
\begin{equation} \label{eq:sumfit}
\overline O_j - \overline B_j = N \sum_k \overline R_{jk} m_k
\, ,
\end{equation}
where $N$ is the APEC normalization of the summed spectrum.
Substituting equations~(\ref{eq:xmmsum}) and (\ref{eq:xmmbgd}) into
equation~(\ref{eq:sumfit}) yields, after some algebra,
\begin{equation} \label{eq:avgrespalmost}
\sum_k \overline R_{jk} m_k = \sum_k m_k \sum_i w^i \frac{N^i}{N} R^i_{jk}
\, ,
\end{equation}
so it is clear that
\begin{equation} \label{eq:avgresp}
\overline R_{jk} = \sum_{i=1}^{10} w^i \frac{N^i}{N} R^i_{jk}
\, .
\end{equation}
The value of the weighted normalization $N$ is given by $N = \sum_i w^i N^i$.

\begin{deluxetable}{lccc}
\tablewidth{0pt}
\tablecaption{{\it XMM} Regions and Spectral Fits \label{tab:reg}}
\tablehead{
&
&
$kT$ &
Norm.\tablenotemark{a} \\
Region &
PIN Weight &
(keV) &
($10^{-2}$ cm$^{-5}$)
}
\startdata
R10 &  0.933\phn &  $8.25\pm0.21$  & $1.67\pm0.01$ \\
R9  &  0.844\phn &  $8.33\pm0.14$  & $4.25\pm0.02$ \\
R8  &  0.746\phn &  $8.27\pm0.15$  & $5.17\pm0.03$ \\
R7  &  0.647\phn &  $8.07\pm0.17$  & $5.09\pm0.03$ \\
R6  &  0.547\phn &  $8.07\pm0.23$  & $4.52\pm0.04$ \\
R5  &  0.448\phn &  $7.40\pm0.34$  & $3.65\pm0.05$ \\
R4  &  0.348\phn &  $7.39\pm0.46$  & $2.93\pm0.05$ \\
R3  &  0.248\phn &  $6.99\pm0.56$  & $2.84\pm0.06$ \\
R2  &  0.147\phn &  $7.65\pm0.44$  & $3.30\pm0.05$ \\
R1  &  0.0421    & $7.45\pm0.68$  & $2.76\pm0.06$ \\
\enddata
\tablenotetext{a}{Normalization of the APEC thermal spectrum,
which is given by $\{ 10^{-14} / [ 4 \pi (1+z)^2 D_A^2 ] \} \, \int n_e n_H
\, dV$, where $z$ is the redshift, $D_A$ is the angular diameter distance,
$n_e$ is the electron density, $n_H$ is the ionized hydrogen density,
and $V$ is the volume of the cluster.}
\end{deluxetable}

For all fits of the {\it XMM} spectra, the energy range considered differed
slightly from the nominal 2-12 keV energy range due to calibration
issues.
At energies near $\sim 2$ keV, there exists a sharp edge in the response
due to gold in the mirrors, and between 8 and 9 keV there are variable background
lines due to copper and zinc.
Neither of these features can be sufficiently accounted for given the current
calibration, and they tend to become important in regions with very good statistics
(all of the Coma {\it XMM} spectra) and/or regions of low surface brightness particularly
near the outer edges of the detector.
We avoid these problems by excluding these features and only
fitting over the range $2.3 < E < 7.5$ keV and $9.5 < E < 12$ keV.
Also, the gain can vary by up to 30 eV, especially during the period
when many of these observations took place (Marcus Kirsch, {\it XMM} Calibration 
Document\footnote{http://xmm2.esac.esa.int/docs/documents/CAL-TN-0018.pdf})
and the redshift found
from fits differs significantly from the nominal value of $z = 0.023$.
We fit the gain with a linear function, assuming the redshift to be 0.0232,
using the {\sc gain} command in {\sc XSPEC} for all 10 spectra.
We adjusted the gain
such that the new response energies $E^\prime$ are related to the original
energies by $E^\prime = E/1.00519 + 0.010312$.
None of these calibration issues have any important effect on the characterization
of the continuum features (IC and hard thermal bremsstrahlung) which are the
subject of this paper.
However, not including these effects would result in high values of $\chi^2$ 
for the fits due to the very
good statistics in the Coma {\it XMM} EPIC-pn spectra, and thus make it more 
difficult to determine the uncertainties in parameters.

Even after these adjustments, fits to the weighted and summed final spectrum
with any model produce high chi-squared values ($\chi^2_{\rm red}\ga1.3$).
A close inspection of the continuum at various energies reveals that the 
residuals are slightly larger relative to the errors than would be expected by
$\chi^2$ statistics, indicating that we have underestimated the errors.
Because of the high signal-to-noise of the Coma observations, the statistical
errors no longer completely dominate over channel-to-channel
systematic effects, caused by, e.g., differing/variable charge transfer 
inefficiencies across the detectors and/or between observations, which were
obtained over a two year period.
We add a 3\% systematic error to the count rate of each channel in the final
spectrum to account for these uncertainties, which leads to more
reasonable values of $\chi^2_{\rm red}\sim1$.

\subsection{{\it XMM} EPIC-pn and {\it Suzaku} HXD-PIN Cross-Calibration}
\label{sec:speccn}

Joint fits between data from different instruments/missions require a
careful consideration of the relative overall calibration if the validity
of fits are to be believed.
Instead of directly finding the cross-normalization through other observations
of a spectrally simple source, such as the Crab, we choose to use the
XIS0 chip of the {\it Suzaku} XRT as an intermediary.
An advantage of this method is that it does not require any assumptions 
about the stability of the
absolute calibration of each instrument between calibration observations
and our observations.
To justify our use of the XIS0 data to calibrate the absolute flux level,
we compare the flux observed by XIS0 to the {\it ROSAT} 0.5-2 keV flux, which
was derived using Snowden's ESAS software package
\citep{SK98}; these fluxes agree to within 1\%.

The overlapping spatial and spectral coverage of the {\it XMM} EPIC-pn and
XIS0 instruments allows a trivial comparison of the flux for a region 
on the sky.
We extract an XIS0 spectrum from the same region as {\it XMM} spectrum R10, 
and we generate rmf and arf files for this region using the 2-7.5 keV
wavelet-smoothed image created from the {\it XMM} EPIC-pn data \citep{SFM+04}.
Though the large XIS PSF ($\sim 2\arcmin$) will scatter photons both into 
and out of this region to a much greater degree than occurs for {\it XMM},
this effect is accounted for in the arf and tied to the {\it XMM} data. 
So while spatial inhomogeneities will not impact the comparison, the
shapes of the spectra will not necessarily be identical.
The overall flux, however, is not sensitive to small variations in the
temperature, and so it provides a good quantity to establish the
{\it XMM}-XIS cross-normalization.
For this region, we find that the 
{\it XMM} flux is 15\% below the XIS0 flux from 2-7.5 keV,
and the XMM flux, extrapolated to 0.5-2 keV, is similarly 15\% below
the {\it ROSAT} flux, and so we scale the summed {\it XMM} arf by this factor.

The cross-normalization between the XIS chips and the HXD-PIN has been
well studied for observations of the Crab nebula
(Ishida et al., {\it Suzaku} Memo 2007-11\footnote{http://www.astro.isas.ac.jp/suzaku/doc/suzakumemo/suzakumemo-2008-03.pdf}).
We adopt their PIN/XIS0 relative normalization factor of $1.132 \pm 0.014$,
increasing the PIN arf, and thus lowering the measured flux, by 13.2\%.

The associated systematic error for both cross-normalization corrections
is estimated to be 1-2\%.
However, the normalization of the R10 
spectrum may differ from that of the other {\it XMM} region spectra, and also the
XIS0-PIN relative normalization may be different due to the fact that 
Coma is spatially
diffuse while the Crab nebula is comparable in size to the XIS resolution.
These issues suggest that
the true cross-normalization systematic uncertainty is probably larger.
We therefore take the combined cross-normalization systematic error to 
be 5\%, which is about as large as can be reasonably allowed by the
simple constraint that a model can be continuously fit across the 12 keV
boundary between the {\it XMM} and PIN spectra.
Specifically, we vary the cross-normalization until the average of the
highest signal-to-noise PIN channels, covering 12 keV $< E < 16$ keV, 
disagrees with the model by $\sim 2-3\sigma$.

\begin{deluxetable*}{lccccc}[b]
\tablewidth{0pt}
\tablecaption{Fits to Joint {\it XMM}-PIN spectra \label{tab:fits}}
\tablehead{
&
$kT$ &
Norm.\tablenotemark{a} &
$\Gamma$ or $kT$ &
& \\
Model &
(keV) &
(cm$^{-5}$) &
(\tablenotemark{b})&
Norm.\tablenotemark{c} &
$\chi^2$/dof
}
\startdata
Single T & $8.45\pm0.06$ & $0.218\pm0.001$ & - & - & 1676.05/1689 \\
T+IC\tablenotemark{d}     & $8.42\pm0.06$ & $0.218\pm0.001$ & -1.6 & $(4.6\pm3.5)\times10^{-9}$ & 1671.29/1688 \\
T+IC\tablenotemark{d}     & $8.45\pm0.07$ & $0.217\pm0.002$ & 2.0 & $(2^{+12}_{-2})\times10^{-4}$ & 1676.18/1688 \\
2T\tablenotemark{e}       & 8.0 & 0.17 & 10.1 & 0.05 & 1672.34/1687 \\
T$_{\rm map}$ & - & - & - & - & 1684.35/1690 \\
\enddata
\tablenotetext{a}{See the note following Table~\ref{tab:reg}.}
\tablenotetext{b}{Value is $\Gamma$ for the T$+$IC model and $kT$ (in keV)
for the 2T model.}
\tablenotetext{c}{Value is the normalization of the power-law component
for the T$+$IC model, which is the photon flux at a photon energy of
1 keV in units of photons cm$^{-2}$ s$^{-1}$ keV$^{-1}$.
For the 2T model, the value is the normalization of the second APEC thermal
model (see the note following Table~\ref{tab:reg}) in units of cm$^{-5}$.}
\tablenotetext{d}{Value of $\Gamma$ is fixed when deriving errors.}
\tablenotetext{e}{Parameters unconstrained.}
\end{deluxetable*}

\subsection{Cosmic X-ray Background} \label{sec:cxb}

We modeled the cosmic X-ray background (CXB) spectrum shape following 
\citet{Bol87}, specifically using the analytical form proposed by
\citet{GMP+99} based on the {\it HEAO-1}/A2+A4 data.
This shape is well-established over the energy range $3<E<60$ keV
and has been confirmed in subsequent measurements
\citep[e.g.][]{RGS+03, Chu+07, Aje+08}.
We adopt a 10\% larger normalization of the spectrum, relative to the original
{\it HEAO-1} determination, to agree with the more
recent measurements by {\it INTEGRAL} \citep{Chu+07}.
This increase is further justified by, and consistent with, 
the ($8\pm3$)\% higher normalization found with {\it Swift} \citep{Aje+08}.
Though these most recent measurements lie slightly, but systematically, 
above the canonical spectrum, as noted by \citet{Aje+08} they are not
inconsistent with other observations at $E>10$ keV.
At the peak of the CXB spectrum, the measurement precision of {\it HEAO-1}
is 10\% \citep{Mar+80}, and the measurement made with the {\it BeppoSAX} PDS
is consistent at the 90\% level with a normalization 12\% larger
\citep{Fro+07}.
In XSPEC, we model the CXB as
\begin{eqnarray} \label{eq:cxb}
{\rm CXB}(E) & = & 1.056 \times 10^{-2} \left(\frac{E}{1 \, 
{\rm keV}}\right)^{-1.29} e^{-E/( 41.13 \, {\rm keV})} 
				\nonumber\\
 & & {\rm photons} \,\, {\rm cm}^{-2} \,\, {\rm s}^{-1} \,\, {\rm keV}^{-1}
\, ,
\end{eqnarray}
where the normalization is set by a $2\arcdeg\times2\arcdeg$ solid angle
of the sky to match the provided response file (see \S~\ref{sec:specpin}).

Cosmic variance due to large scale structure depends on the solid angle
of the observation ($\Omega = 0.32$ deg$^2$ for the PIN) and on the cut-off flux
of removed point sources ($S_{\rm cut}$), determined by the {\it XMM} source 
completeness \citep{FBH+04} to be
$S_{\rm cut}(12-70 \, {\rm keV}) = 2.2\times10^{-13}$ erg s$^{-1}$ cm$^{-2}$.
Since the variance 
$\sigma_{\rm CXB}/I_{\rm CXB} \propto \Omega^{-0.5} S_{\rm cut}^{0.25}$,
we can estimate the variance in our observation relative to another 
measurement assuming a $\log N$--$\log S$ relation of $N(S) \propto S^{-1.5}$.
Using the HEAO-1 A2 estimate \citep{Sha83,BMC00,RGS+03}
with $\Omega = 15.8 \, {\rm deg}^2$,
$S_{\rm cut} = 8\times10^{-11}$ erg s$^{-1}$ cm$^{-2}$, and
$\sigma_{CXB}/I_{CXB} = 2.8\%$ (1$\sigma$), we find a variance of
7.4\% (90\% confidence), which we take as an additional systematic 
error in the PIN flux.
To account for the 10\% discrepancy between the HEAO-1 and the 
{\it INTEGRAL} \citep{Chu+07} and {\it Swift} \citep{Aje+08}
observations, we also estimate the standard deviation of these
measurements to be 7\% (90\% confidence).
Adding these uncertainties in quadrature, we take the total systematic error 
in the CXB normalization to be 10\%.
Below 20 keV, the CXB emission is $\la 10$\% of the total flux, and it
just becomes the dominant source of emission at $\sim 50$ keV.

For the {\it XMM} data, the background spectra include unresolved point
sources that make up the CXB, so they do not need
to be modeled.

\subsection{Point Sources} \label{sec:ptsrcs}

Point sources in the {\it XMM-Newton} Coma mosaic have already been identified
by \citet{FBH+04}, who also give their count rates in three
energy bands ($0.5-1$ keV, $1-2$ keV, and $2-4.5$ keV).
For each of the 72 sources, we assume the spectrum to be described by
an unabsorbed power law and fit this model to each spectrum.
We found that the sum of all these models, weighted by $w^i$
according to their positions, could be more concisely described by the
sum of two power laws with photon indices 2.1 and 1.6 and normalizations
$8.54\times10^{-5}$ and $1.23\times10^{-4}$ photons cm$^{-2}$ s$^{-1}$ keV$^{-1}$
at 1 keV, respectively.
While a simple power law description poorly characterizes some of the
sources, care is taken to ensure that individual fits, when extrapolated to high
energies, are not unphysical.
Their composite spectrum accounts for $\la 1$\% of the {\it XMM}
flux and is therefore unimportant relative to other systematic effects.
For this reason we do not go to the extra effort to exclude the sources 
from the {\it XMM} spectra.
Assuming the spectral fits are reasonably valid, the point sources
account for $\sim 10\%$ of the CXB at PIN energies.
During fits of the joint {\it XMM} and {\it Suzaku} data, we include this point
source composite model for both spectra.

The brightest of these point sources is X Comae, a background AGN with a
flux $\sim10\times$ brighter than any other source in the field.
From {\it XMM} RGS observations, it is known to have a steep spectrum
($\Gamma\sim2.4$) and to vary in flux by about a factor of 2 over the
course of 1 year \citep{THF+07}.
However, due to its position, nearly 90\% of the
flux from X Comae is not detected by the HXD, so this source does not
significantly contribute to the PIN spectrum.

\section{Spectral Fits} \label{sec:fits}

In our spectral fits, all model components are absorbed by the neutral 
hydrogen column density toward Coma, $N_H = 9 \times 10^{19} {\rm cm}^2$
[average of values derived from \citet{DL90} and \citet{KBH+05}],
though this absorption is negligible at energies above 2 keV.
In general, we characterize the dominant thermal emission in XSPEC
with the APEC model for $E<40$ keV and with the MeKa model for $E>40$ keV.
In the currect version of XSPEC, the APEC and MeKaL models are undefined
above 50 keV, though the MeKa and bremsstrahlung models are defined.
We tie the parameters of the MeKa model to the APEC parameters, except for
the MeKa normalization, which we reduce relative to the APEC normalization by
5.5\% to bring the models into agreement at high energies.
Also, the abundances relative to solar and the redshift are fixed, to
0.24 and 0.0232 respectively (see \S~\ref{sec:specxmm}).
This value for the abundance is based on fits to the final weighted and 
summed {\it XMM} spectrum alone, and the best-fit abundances of all the 
individual spectra from the 10 regions is also consistent with this value.
Since we are interested in continuum features, the exact choice for
the abundance does not strongly affect the results.
The spectral fitting results are summarized in Table~\ref{tab:fits}.

\subsection{Joint {\it XMM-Newton} and {\it Suzaku} Spectral Fits 
Without Considering Systematic Errors}\label{sec:fitsjoint}

\begin{figure}
\plotone{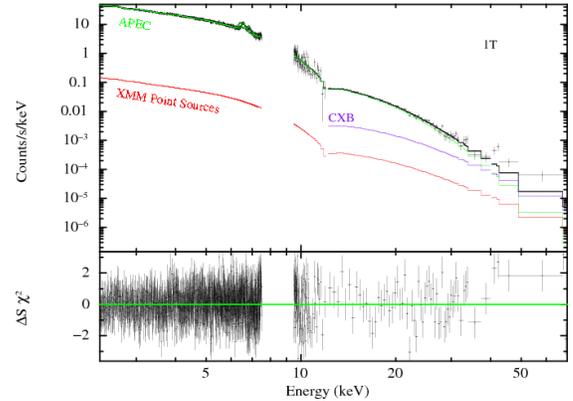}
\caption{{\it Suzaku} HXD-PIN spectrum ($E > 12$ keV) and
the combined {\it XMM} spectrum ($E < 12$ keV) corresponding to
the spatial sensitivity of the PIN. 
Shown as solid lines are the best fit models for a single temperature
thermal component.
The thermal model (``APEC", green) is nearly coincident with the data, 
though falling below it at higher energies.
Also included for all joint fits are the the total spectrum for the
``{\it XMM} Point Sources" (red) and the Cosmic X-ray Background (``CXB", purple),
the latter of which only applies to the PIN spectrum since the CXB
is subtracted from the {\it XMM} data along with the NXB.
\label{fig:spec1T}}
\end{figure}

We simultaneously fit the {\it Suzaku} HXD-PIN and {\it XMM-Newton} 
EPIC-pn spectra for the PIN FOV.
First, we consider only a single temperature fit, in order to establish
whether the addition of a non-thermal component actually improves the fit
(Fig.~~\ref{fig:spec1T}).
We find a best fit temperature of $8.45 \pm 0.06$ keV,
which is in general
agreement with similar fits to the PIN data ($8.3 \pm 0.3$ keV) and
{\it XMM} data ($8.37 \pm 0.12$ keV) individually.
Note that the dip at 15 keV is a known problem with the current NXB model
(Mizuno et al., {\it Suzaku} Memo
2008-03\footnote{http://www.astro.isas.ac.jp/suzaku/doc/suzakumemo/suzakumemo-2008-03.pdf}).
Since each spectrum is individually described by the same average
temperature, the existence of excess emission at hard energies is unlikely.
While all of these temperatures are slightly higher than the
cluster-wide average temperature of 8.2 keV \citep{HBS+93},
the energy range in this and similar fits typically extends to energies below 2 keV 
and thus includes more low temperature gas.

\begin{figure}
\plotone{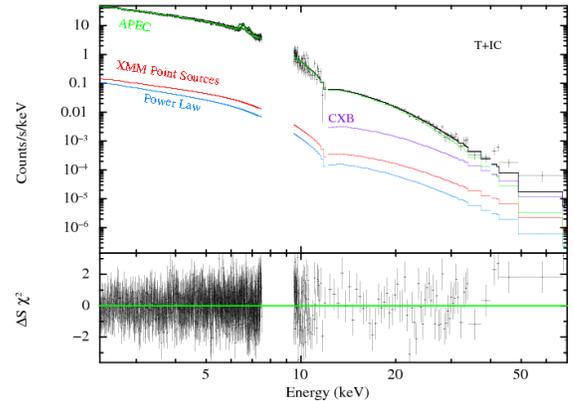}
\caption{{\it Suzaku} HXD-PIN spectrum ($E > 12$ keV) and
the combined {\it XMM} spectrum ($E < 12$ keV) corresponding to
the spatial sensitivity of the PIN. 
Shown as solid lines are the best fit models for a single temperature
thermal component plus a non-thermal component.
The thermal model (``APEC", green) is nearly coincident with the data, 
though falling below it at higher energies.
The non-thermal model (``Power Law", light blue) is the faintest model
component for both spectra, and the photon index is fixed at $\Gamma=2.0$.
The other two components are described in Figure~\ref{fig:spec1T}.
\label{fig:specTIC}}
\end{figure}

The addition of a power-law non-thermal component produces a formally better 
description of the spectra, according to the f-test, improving the
overall fit (Table~\ref{tab:fits}), but {\it only} for a photon index $\Gamma<0$.
Allowing the temperature and power law photon index to vary along with each
component's normalization, we find $T = 8.42 \pm 0.06$ keV and
$\Gamma = -1.6$, though $\Gamma$ is poorly constrained. 
If we fix $\Gamma$ to this best-fit value, the IC component is significant
at the $2.2\sigma$ level without considering the effect of systematic
uncertainties.
However, this photon index is completely inconsistent with the spectral
index of the radio halo
\citep[$\Gamma \ga 1.8$,][]{GFV+93}.
While we might expect a flatter spectrum for IC emission,
since the hard X-ray photons are emitted by somewhat lower energy electrons
than the radio emission, and the radio spectrum flattens at 
lower frequencies \citep{TKW03},
a rising IC spectrum with energy is completely unexpected and unphysical.
The power law fit, in contrast to finding an actual power law signature
in the data, is instead compensating for a slight excess at high energies
while minimizing its impact on the overall fit at lower energies.
Notice that the residuals in Figure~\ref{fig:spec1T} above 40 keV lie
systematically, if not significantly, above the model.
This excess at energies above 40 keV can be explained as a $\sim2$\% underestimate
of the NXB, as suggested by the Earth-occulted spectrum 
(see \S~\ref{sec:specpin}).
Increasing the background level by 2\% for $E>40$ keV results in a best-fit
power law component very similar to the model used for the {\it XMM} point sources,
with $\Gamma\sim1.6$, but it is not significant at even the $1\sigma$ level.
A similar result is found if the photon index is fixed at $\Gamma=2$ and
the NXB above 40 keV is not increased; this fit is shown in 
Figure~\ref{fig:specTIC}.
In this case, the fit is not improved by the addition of a power-law component
to the model.

\begin{figure}
\plotone{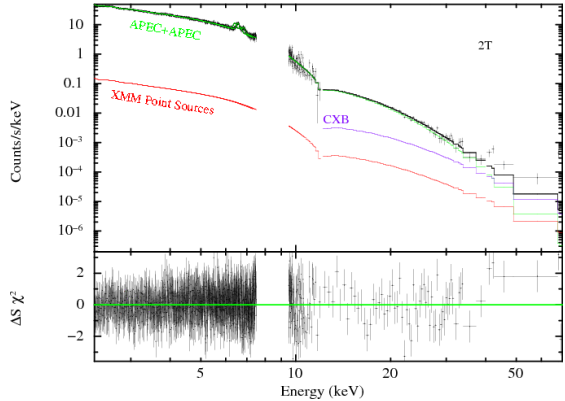}
\caption{{\it Suzaku} HXD-PIN spectrum ($E > 12$ keV) and
the combined {\it XMM} spectrum ($E < 12$ keV) corresponding to
the spatial sensitivity of the PIN. 
Shown as solid lines are the best fit models for a two-temperature
thermal component.
The thermal model (``APEC+APEC", green) is nearly coincident with the data, 
though falling below it at higher energies.
The other two components are described in Figure~\ref{fig:spec1T}.
\label{fig:spec2T}}
\end{figure}

Interestingly, a two-temperature model for the ICM yields only a slightly
better fit to the data than does the single temperature model
(see Fig.~\ref{fig:spec2T}), though the
addition of a second temperature component is probably not formally justified.
This result is mainly due to the addition of the 3\% systematic error to
the {\it XMM} spectrum.
Without including that error, a two-temperature model produces a 
clearly improved fit over a single temperature model, indicating that
the addition of this error is somewhat obscuring evidence for a
multi-temperature continuum.
In either case, the two temperatures are not strongly constrained, but they are 
broadly consistent with the spatial variations in Coma's temperature
(see \S~\ref{sec:tmap} and Fig.~\ref{fig:tmap}).
Therefore, even before systematic errors are considered, the case for the
inclusion of a non-thermal component is not strongly motivated.

\subsection{Multiple Thermal Components in Coma} \label{sec:tmap}

In most previous attempts to measure a non-thermal component in the hard spectrum
of Coma, the thermal emission was modeled as a single temperature plasma
characteristic of the average global state of the ICM.
However, Coma is known to host temperature variations \citep[e.g.][]{HHE+98}.
Generally, all clusters exhibit a multi-temperature ICM \citep{CDM+08}, and
this is especially true of merging clusters like Coma,
which tend to host hot regions due to shocks \citep{MFS+98}.
At hard energies, where the exponential turnover in the bremsstrahlung
continuum is especially well sampled, even weak higher temperature
components can significantly contribute to the flux.
Also, these components would lead to a higher average temperature for the
ICM than if the average cluster temperature were determined from the
spectrum at softer energies, such as from 0.5-10 keV.

\begin{figure}
\plotone{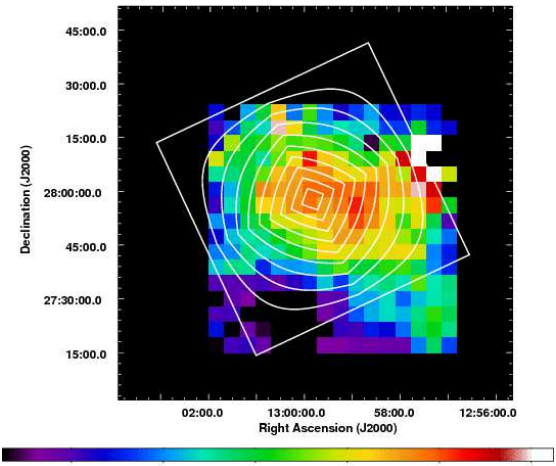}
\caption{{\it XMM-Newton} temperature map across Coma with HXD-PIN contours of
constant PIN effective area overlaid at 10\% intervals.
The {\it XMM-Newton} spectra were fit in square spatial regions 4\farcm3 on a side.
The temperatures, given in keV by the color bar, are accurate to either a
few tenths of a keV (in the center) or 1--2 keV in lower surface brightness
regions.
Temperatures shown here were determined from fits to the 0.5--14 keV 
spectrum in each region.
\label{fig:tmap}}
\end{figure}

In the previous section, we found that a two temperature model provided
a slightly better description than did a single temperature model of
joint fits to the {\it XMM-Newton} and {\it Suzaku} data, especially
when ignoring the 3\% systematic error applied to the {\it XMM} data.
This may indicate that there are multiple temperature components in Coma.
The multiple components could occur along the line-of-sight, or in the plane
of the sky, or locally (the gas might be multiphase).
In fact, previous temperature maps show that Coma certainly has temperature 
structure which is likely associated with mergers \citep{WYF+99}.
Here, we test whether this temperature structure alone could reproduce the
observed {\it Suzaku} PIN spectrum of Coma, without any non-thermal emission.
From the {\it XMM-Newton} EPIC-pn mosaic of Coma, we constructed a temperature
map on a $16\times16$ grid with cell size of $4\farcm3$ on a side.
Each of the spectra were fit with a single temperature APEC model to produce a
temperature map that covers the Coma mosaic, as shown in Figure~\ref{fig:tmap}.
We weighted these model fits by the PIN spatial sensitivity and 
combined them.
This resulting model was compared to the PIN spectrum (Table~\ref{tab:fits},
row labeled ``T$_{\rm map}$").
Note that only the overall normalization of the T$_{\rm map}$ model was
allowed to be fit, to compensate for a loss of flux due to incomplete
coverage of the map across the HXD FOV.
Also, the spectral shape and normalization of 
each of the thermal models was the same as given by the {\it XMM-Newton}
temperature, and each model was simply weighted by the average PIN effective
area at that position.

\begin{figure}
\plotone{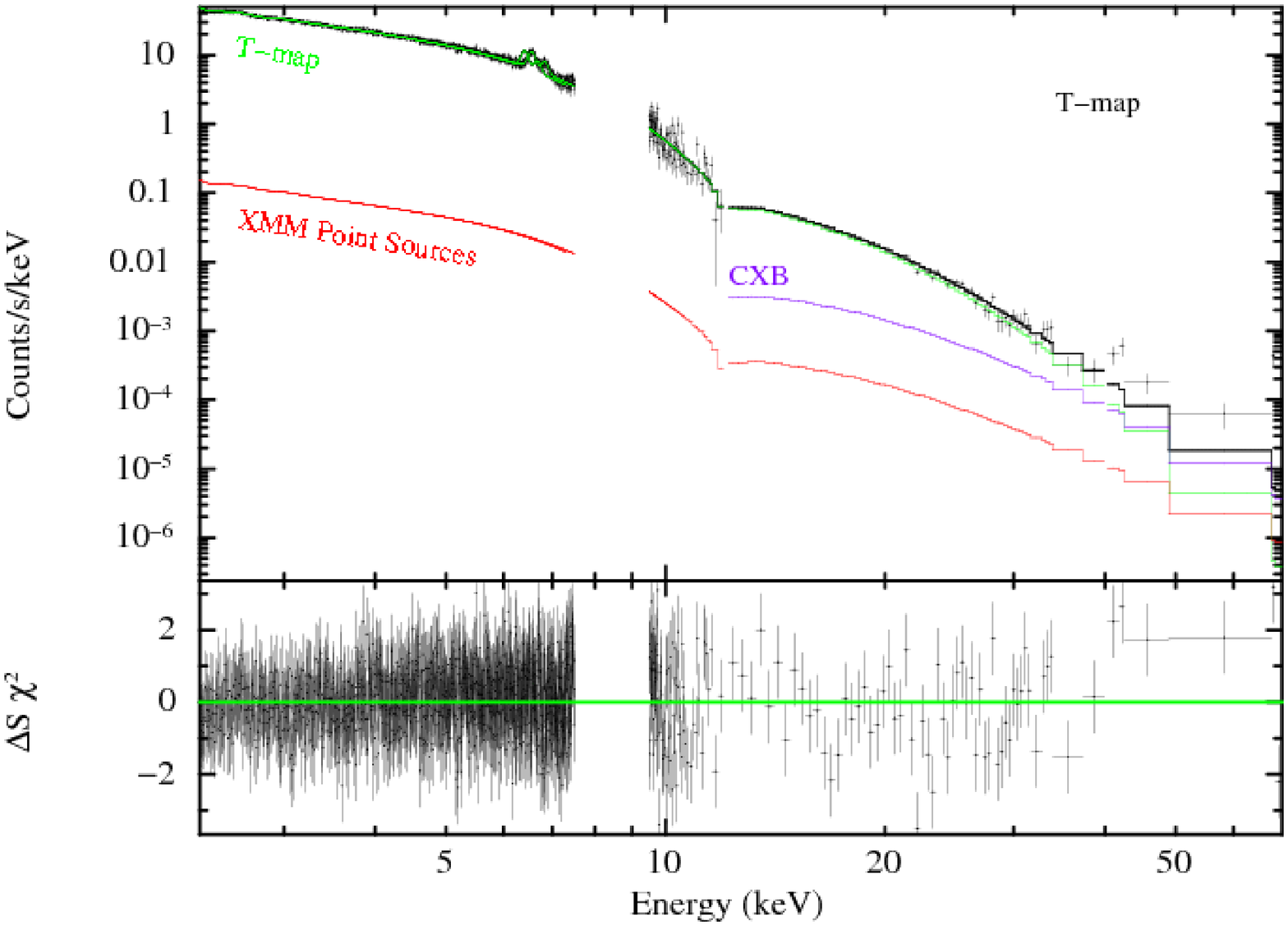}
\caption{{\it Suzaku} HXD-PIN spectrum ($E > 12$ keV) and
the combined {\it XMM} spectrum ($E < 12$ keV) corresponding to
the spatial sensitivity of the PIN. 
Shown as solid lines are the combined spectra of the best fit models from
the temperature map.
The thermal model (``Tmap", green) is nearly coincident with the data, 
though falling below it at higher energies.
The other two components are described in Figure~\ref{fig:spec1T}.
\label{fig:specTmap}}
\end{figure}

This T$_{\rm map}$ model provides a good fit to the PIN spectrum 
with no adjustable parameters that affect the spectral shape
(Fig.~\ref{fig:specTmap}).
While the value of $\chi^2$ is worse than for all previous fits, it is only 
slightly worse, which is not unexpected since it is not really a fit at all.
The fact that {\it XMM}-derived temperatures extrapolated to PIN energies are
sufficient to fully account for the PIN spectrum further suggests that 
no non-thermal hard X-ray excess has been detected with the PIN,
especially below 40 keV and probably below 70 keV.
Also, simulating spectra of similar quality to our data, assuming the
T$_{\rm map}$ model for the underlying source, yields a joint single 
temperature fit consistent
with that found from the actual data, with $T=8.51\pm0.06$ keV.

\subsection{Systematic Errors in the Spectrum} \label{sec:syst}

\begin{figure}
\plotone{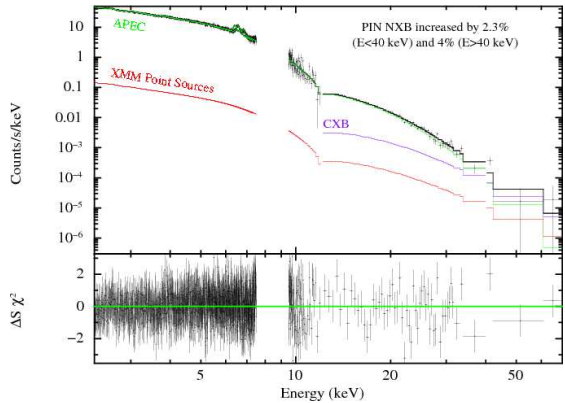}
\caption{{\it Suzaku} HXD-PIN spectrum ($E > 12$ keV) and
the combined {\it XMM} spectrum ($E < 12$ keV) corresponding to
the spatial sensitivity of the PIN. 
The PIN NXB is increased to its 90\% confidence limit,
which is 2.3\% for $E<40$ keV and 4\% for $E>40$ keV.
The thermal model (``APEC", green) is nearly coincident with the data, 
though falling below it at higher energies.
The other two components are described in Figure~\ref{fig:spec1T}.
Note that the residuals above the model that exist in the previous fits
at $E>40$ keV have disappeared.
\label{fig:specnotul}}
\end{figure}

We explicitly consider the systematic error for 3 quantities: the PIN
non-X-ray background, the {\it XMM}-{\it Suzaku} cross-normalization factor, 
and the normalization of the CXB as modeled for the PIN spectrum.
To test the effect of these systematic errors, we vary the relevant quantity
up and down by our estimate of the 90\% systematic error, and evaluate the
resulting change in best-fit model parameters.
The detection of a non-thermal component cannot be claimed unless it remains 
robust to variations of these quantities within their systematic errors.
Because the largest error is in the normalization of the PIN NXB, we first
increase it by 2.3\% from 12-40 keV and 4\% from 40-70 keV and repeat the single
temperature plus non-thermal model fit.
The new best-fit IC normalization is pushed to zero. 
Even before considering the effect of other systematic errors, from this
exercise alone it is clear that we {\it do not detect} non-thermal 
emission in the HXD-PIN spectrum, given the current uncertainty in the NXB
normalization.
This fit, with a temperature for the thermal component of $8.33\pm0.06$ keV,
is shown in Figure~\ref{fig:specnotul}.
In fact, the value of $\chi^2$ is slightly
lower ($\chi^2=1672.25$ for 1689 dof) than for the nominal PIN NXB single
temperature fit.
Notice that the residuals above the model for $E>40$ keV, seen in all the
previous spectral fits, have disappeared.

Though we cannot claim to detect non-thermal emission, we can derive an
upper limit to its flux based on joint fits to the spectra, including
systematic errors in the following way.
First, for an assumed photon index which we fix, we find the nominal
normalization $N_{\rm nom}$ and corresponding 90\% upper bound 
$N_{\rm nom}^{\rm ul}$ of the non-thermal component for a single temperature 
plus power law model, allowing the temperature and normalization of the 
thermal component to vary.
Then, for each systematically uncertain quantity, we set that quantity to
the limit bounded by the systematic error in the sense that increases
the value of the non-thermal normalization $N_{{\rm sys},i}$, and we
fit for it and its 90\% upper bound $N_{{\rm sys},i}^{\rm ul}$.
The statistical and systematic errors of the power law normalization
are then given by
\begin{equation} \label{eq:delstat}
\delta_{\rm stat} = N_{\rm nom}^{\rm ul} - N_{\rm nom} \, ,
\end{equation}
and
\begin{equation} \label{eq:delsys}
\delta_{{\rm sys},i} = N_{{\rm sys},i}^{\rm ul} - N_{\rm nom}^{\rm ul} =
N_{{\rm sys},i} - N_{\rm nom}
\, ,
\end{equation}
respectively.
The final 90\% upper limit is then given by
\begin{equation} \label{eq:ul}
N^{\rm ul}_{\rm tot} = N_{\rm nom} + \sqrt{\sum_i \delta_{{\rm sys},i}^2
+ \delta_{\rm stat}^2}
\, .
\end{equation}
We add each systematic error contribution in quadrature because it is
unlikely that we chose normalizations for these 3 quantities such that each
one disfavors the detection of non-thermal emission in the most severely
possible way.

\begin{figure}
\plotone{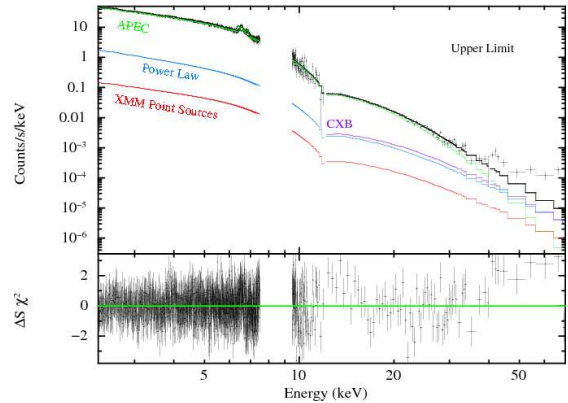}
\caption{{\it Suzaku} HXD-PIN spectrum ($E > 12$ keV) and
the combined {\it XMM} spectrum ($E < 12$ keV) corresponding to
the spatial sensitivity of the PIN. 
All quantities with systematic uncertainties (PIN CXB and NXB, the
{\it XMM}-{\it Suzaku} cross calibration) are set to their 90\% confidence limit
in the direction that favors the addition of a power law model component
to describe the data.
The thermal model (``APEC", green) is nearly coincident with the data, 
though falling below it at higher energies.
The non-thermal model (``Power Law", light blue) is shown for $\Gamma=2$
at its 90\% confidence upper limit value.
The other two components are described in Figure~\ref{fig:spec1T}.
\label{fig:specul}}
\end{figure}

The upper limits for a range of assumed photon indices is provided in
Table~\ref{tab:uls}, and in Figure~\ref{fig:specul} we show, for $\Gamma=2$,
the resulting best fit with all 3 systematic quantities set at the limit
of their 90\% confidence range.
Over the PIN energy band (12-70 keV), the flux is relatively independent 
of photon index.
To compare our results to the most recent previous detections of non-thermal
emission in Coma, we also give the upper limit on the non-thermal flux
in the 20-80 keV band, which is $6.0 \times 10^{-12}$ erg s$^{-1}$ cm$^{-2}$
for $\Gamma = 2$.
This limit is inconsistent with the {\it RXTE} \citep{RG02} and 
{\it BeppoSAX} \citep{FOB+04}
detections by about a factor of 2.5, but at the same level as the upper limit
derived by \citet{RM04} from the {\it BeppoSAX} data.
If we adopt a 4\% systematic error for the PIN NXB instead of 2.3\% 
for $E<40$ keV, which
would better agree with that derived from ``blank sky" observations,
then our upper limit increases by 35\%.
Similarly, if we also increase the CXB and {\it XMM}-{\it Suzaku}
cross-normalization to 18\% and 10\%, respectively, our upper limit for
$\Gamma=2$ would increase by 50\%.
In either case, our upper limit still excludes the {\it RXTE} and
{\it BeppoSAX} detections, if FOV differences are ignored (see 
\S~\ref{sec:disc} for a more meaningful comparison).

\begin{deluxetable*}{ccccc}
\tablewidth{0pt}
\tablecaption{90\% Upper Limits on IC Flux \label{tab:uls}}
\tablehead{
&
Norm.\tablenotemark{a} &
Flux (12-70 keV) &
Flux (20-80 keV) &
{\it BeppoSAX} Detection\tablenotemark{b} \\
$\Gamma$ &
($10^{-3}$ photons keV$^{-1}$ cm$^{-2}$ s$^{-1}$) & 
($10^{-12}$ erg s$^{-1}$ cm$^{-2}$) &
($10^{-12}$ erg s$^{-1}$ cm$^{-2}$) &
($10^{-12}$ erg s$^{-1}$ cm$^{-2}$)
}
\startdata
1.0 & 0.155    & 14.4\phn\phn & 14.9\phn\phn & \\
1.1 & 0.220    & 14.2\phn\phn & 14.4\phn\phn & \\
1.2 & 0.311    & 14.1\phn\phn & 13.9\phn\phn & \\
1.3 & 0.439    & 13.9\phn\phn & 13.4\phn\phn & \\
1.4 & 0.617    & 13.8\phn\phn & 12.9\phn\phn & \\
1.5 & 0.860    & 13.5\phn\phn & 12.3\phn\phn & \\
1.6 & 1.18\phn & 13.1\phn\phn & 11.6\phn\phn & \\
1.7 & 1.58\phn & 12.4\phn\phn & 10.7\phn\phn & \\
1.8 & 2.04\phn & 11.4\phn\phn & 9.51 & \\
1.9 & 2.48\phn & 9.83         & 7.98 & \\
2.0 & 2.70\phn & 7.64         & 6.01 & $15\pm5$ \\
2.1 & 2.70\phn & 5.46         & 4.15 & \\
2.2 & 2.56\phn & 3.71         & 2.73 & \\
2.3 & 2.36\phn & 2.46         & 1.74 & \\
2.4 & 2.15\phn & 1.62         & 1.11 & \\
\enddata
\tablenotetext{a}{Normalization of the power-law component
for the T$+$IC model, which is the photon flux at a photon energy of
1 keV.}
\tablenotetext{b}{Flux (20-80 keV), as reported in \citet{FOB+04}.}
\end{deluxetable*}

\section{Implications and Discussion} \label{sec:disc}

After modeling all the known possible contributions to the $2-70$ keV 
spectrum, simultaneously fitting for the parameters of thermal and
non-thermal spectral components, and taking into account the systematic
uncertainty of the PIN NXB,
we do not see evidence for IC emission in Coma at our level
of sensitivity.
We therefore derive an upper limit to non-thermal, hard X-ray emission
through a careful consideration of the maximum effect
of systematic uncertainties on our ability to detect a non-thermal
signal.
This conservative upper limit is similar to that derived by \citet{RM04}
and is inconsistent with claimed detections using {\it RXTE} \citep{RG02}
and {\it BeppoSAX} \citep{FOB+04} by approximately a factor of 2.5.
However, it should be noted that we do not include potentially 
lost emission due to PIN vignetting from any of our flux upper limits
relative to the larger FOVs of {\it RXTE} and {\it BeppoSAX}, which have
collimator FWHM of $1\arcdeg$ and $1.3\arcdeg$, respectively.

If IC emission follows the radio synchrotron emission [as derived from the
point source-subtracted radio image from \citet{DRL+97}],
as it would for a 
uniform $B$ field throughout the cluster, our upper limits imply a total IC
flux $1.7-2\times$ larger would be found inside an 
{\it RXTE}/{\it BeppoSAX}-like FOV.
We also consider a more detailed spatial distribution for the underlying IC
emission, derived from the re-acceleration model of \citet{BB05}, in which
relativistic protons collide with electrons in the ICM that are then
re-accelerated by Alfv\'en waves due to cluster mergers.
Given the radio spectrum of Coma, this model predicts that the smaller FOV
of the {\it Suzaku} HXD-PIN would lead to an underestimate of the
non-thermal flux by a factor of $2-2.5$ (possibly 3 under extreme
circumstances).
If this model for the spatial distribution of the non-thermal emission is
correct, then our upper limit is just
consistent with the measurements of \citet{RG02} and \citet{FOB+04}.
(However, the \citet{BB05} model actually predicts a non-thermal flux
considerably below the {\it BeppoSAX} and {\it RXTE} detections.)
Because any spatial variation of the magnetic field strength is unknown,
a direct comparison between these missions is not possible with any precision.
Under the reasonable assumption that $B$ decreases with radius, our
upper limit will be $\ga2\times$ larger, so our result cannot definitively
rule out the detections discussed above.
Regardless of this issue, the upper limit is approximately the same as
or slightly higher than the upper limit range found by \citet{RM04}.

However, the {\it BeppoSAX} PDS measures a 20-80 keV flux for the Crab
of $1.23\times10^{-8}$ erg s$^{-1}$ cm$^{-2}$ \citep{Kir+05}, 
while the {\it Suzaku} PIN flux over this energy range is 
$1.56\times10^{-8}$ erg s$^{-1}$ cm$^{-2}$, after applying the 13.2\% correction
to bring the PIN spectrum into agreement with the XIS fluxes 
(Ishida et al., {\it Suzaku} Memo 2007-11\footnote{http://www.astro.isas.ac.jp/suzaku/doc/suzakumemo/suzakumemo-2007-11.pdf}).
This 21\% flux difference implies our upper limit would be 
$4.7\times10^{-12}$ erg s$^{-1}$ cm$^{-2}$ on the {\it BeppoSAX} scale, which is
on the lower end of the range estimated by \citet{RM04}.
Also, even if we only detect one-third of the total emission observed by
the {\it BeppoSAX} PDS, we would just barely exclude the nominal value of the
\citet{FOB+04} measurement.

Assuming the differing measurements of non-thermal emission are not
due to the IC radiation having a larger extent, what might be the cause
of this discrepency?
While it could be explained by a greater point source
contamination at hard energies for the {\it RXTE} and {\it BeppoSAX} missions
due to their larger FOVs, most likely we differ in our results
because of different considerations of the thermal gas.
For both detections, the gas temperature was found to be 
lower than our nominal value of 8.45 keV.
Fixing the gas temperature to their assumed values in our fits yields
a $\Gamma=2.0$ non-thermal component significance $>4\sigma$, without
including systematic effects, for $T=7.67$ keV ({\it RXTE}) and 
$T=8.2$ keV ({\it BeppoSAX}); however, these fits are poor relative to 
fits in which the temperature is a free parameter.
Though these temperatures differ from our best-fit value by only
a few percent, the exponential decline of bremsstrahlung continuum at high
energies amplifies even small differences.
The lower measurements of the ICM temperature appear not to be due to
the inclusion of data at low energies ($E<1$ keV), which can bias average
temperature estimates low.
Most likely, the larger FOVs of {\it RXTE} and {\it BeppoSAX} allowed the
inclusion of emission from more cool gas in the cluster outskirts than
was observed by {\it Suzaku}.
This emission would serve to lower the average observed temperature, which
is primarily determined from emission at lower energies ($E<10$ keV).
But, as evidenced by the temperature map in Figure~\ref{fig:tmap}, a
distribution of higher-than-average temperature regions can effectively
increase the average gas temperature observed at high energies,
as first seen by \citet{NLP+03}.

We take the reasonably good agreement between the thermal models
derived from the {\it XMM} temperature map and the PIN spectrum to mean
that we essentially only detect thermal emission from Coma out to
70 keV.
This result is fully consistent with recent {\it INTEGRAL} detections of 
extended hard X-ray emission.
\citet{RBP+06} performed a point-by-point spectral comparison between
{\it XMM}-derived and {\it INTEGRAL}-derived temperatures and found that they
followed a strict one-to-one correlation.
Similarly, \citet{ENC+07} characterized a surface brightness excess relative
to the {\it XMM} data, which they found to be best described by extended
hot, thermal emission at a $T \sim 12 \pm 2$ keV.
This excess coincides with the hotter temperatures ($T \sim 10-11$ keV)
to the west of the PIN pointing center in Figure~\ref{fig:tmap}.

From our upper limit on the flux of IC emission, we can derive a lower
limit on the average magnetic field strength $B$ as shown by 
\citet{HR74}.
Equation~(\ref{eq:synicratio}) refers to the total energy emitted for one
electron; it is more useful to consider the ratio of monochromatic
fluxes $F_R(\nu_R)$, $F_X(\nu_X)$ at frequencies $\nu_R$, $\nu_X$, 
for a power law distribution of electrons, from which
we can derive an expression for the magnetic field as
\begin{equation} \label{eq:bexact}
B = C(p) (1+z)^{(p+5)/(p+1)} \left(\frac{F_R}{F_X}\right)^{2/(p+1)}
   \left(\frac{\nu_R}{\nu_X}\right)^{(p-1)/(p+1)}
\, ,
\end{equation}
where $p$ is the index of the electron distribution $N(E) \propto E^{-p}$
and is related to the spectral index $\alpha$ ($F_\nu \propto \nu^{-\alpha}$)
by $p=2\alpha+1$.
The value of the proportionality constant $C(p)$ can be found from
the ratio of the synchrotron flux \citep[][eqn.~18.49]{Lon94} to the
IC flux \citep[][eqn.~7.31]{RL79}.
Assuming that the electron energy
distribution does not turn over significantly at low energies
and that $\alpha=1$, we find $B>0.15$ $\mu$G.
This limit is still below the equipartition value of 0.5 $\mu$G 
\citep{GFV+93},
and it is well below the line-of-sight estimates of a few $\mu$G
derived from Faraday rotation measure (RM) studies \citep{FDG+95}.
Note that the Faraday RM magnetic field estimates are sensitive to the
$B$ field geometry and may imply a field strength larger than the
volume-averaged value if $B$ is preferentially aligned along filamentary
structures on small scales \citep{Pet01}.
Also, we are unable to put interesting constraints on the relativistic
energy budget of Coma, since our lower limit includes the equipartition 
estimate of $B$, which defines the minimum energy in relativistic
components and would not imply a significant contribution
to the energy budget of Coma.

It has been noted that the hard excess detected by \citet{ENC+07} also corresponds
to the peak in the point source-subtracted image from \citet{DRL+97},
potentially indicating that the hard emission could in fact be 
non-thermal in origin \citep{EPN+07}.
In fact, we suspect that this peak, which appears tantalizingly like
a small radio relic, is not a true feature of the halo, but instead is
the result of imperfect source subtraction.
Due to the large beam size used to create the diffuse radio image,
extended emission from radio galaxies might not have been properly subtracted
using a point source list.
We point out that the strongest radio source (1256+282 or 5C 4.81,
centered on NGC 4869)
in Coma is near this position, is a head-tail radio galaxy with a
steeper spectral index than rest of the halo \citep{GFV+93}, and that its
tail extends to the west \citep[][see Fig.~1(j)]{OO85} and turns north 
\citep[][see Fig.~2]{VGF90}.
Subtracting a point source from this morphology would leave a residual
very similar to that in the Deiss image.
Therefore, any relation between the location of hard emission and this
radio feature probably should be regarded as coincidence.

As the calibration of the NXB model improves, constraints on IC emission
in Coma will tighten, possibly leading to a detection.
The uncertainty in the current background model, ``bgd\_d," is more than
a factor of 2 lower than the original model.
However, the existence of non-thermal emission in the Coma cluster may
have to be determined by future missions --- particularly, those missions
with hard X-ray imaging capabilities like
NuStar\footnote{http://www.nustar.caltech.edu/},
Astro-H (previously NeXT)\footnote{http://www.astro.isas.ac.jp/future/NeXT/},
and Simbol-X\footnote{http://www.asdc.asi.it/simbol-x/}.
If the IC emission is localized, then our joint fitting-technique can
be used for many much smaller regions where temperature-mixing will be
less significant and the IC component will be relatively stronger.
Also, the $B$ field strength can be derived spatially across a cluster,
yielding a better estimate of the possible pressure support provided
by relativistic components in the ICM, which could modify mass
estimates that depend on the hydrostatic equilibrium state of the cluster gas.

\acknowledgments
We thank W. Reich who kindly provided us with the
Deiss et al.\ (1997) radio image.
We thank
M. Ajello,
G. Brunetti,
F. Fusco-Femiano,
S. Molendi,
and
P. Rebusco
for useful discussions. 
Also, we are grateful to the referee for helpful comments that 
improved the paper.
Support for this work was provided by
NASA {\it Suzaku} grants NNX06AI44G and NNX06AI37G and
{\it XMM-Newton} grant NNX06AE76G.
DRW was supported by a Dupont Fellowship and a Virginia Space Grant 
Consortium Fellowship.
AF was partially supported by NASA NNG05GM5OG grant to UMBC.
Basic research in radio astronomy
at the NRL is supported by 6.1 Base funding.

\end{document}